\documentclass{webofc}
\usepackage[varg]{txfonts}   
\usepackage{MnSymbol}
\usepackage[utf8]{inputenc}
\usepackage{soulutf8}
\usepackage{graphicx}
\usepackage{epstopdf}
\usepackage{amsmath}
\usepackage{slashed}
\usepackage{xcolor}
\usepackage{mathtools}
\usepackage{slashed}

\usepackage{array,multirow}
\usepackage{booktabs}
\usepackage{adjustbox} 
\usepackage{rotating}
\usepackage{blindtext}
\usepackage{physics}

\begin{document}
\title{Two Schr\"odinger-like Equations for hadrons}
%
%

\author{\firstname{Ruben} \lastname{Sandapen}\inst{1}\fnsep\thanks{\email{ruben.sandapen@acadiau.ca}} }

\institute{Department of Physics, Acadia University, Wolfville, Nova-Scotia, Canada, B4P 2R6}

\abstract{In this talk, based on \cite{Ahmady:2021yzh,Ahmady:2021lsh}, I argue that the holographic Schr\"odinger Equation of $(3+1)$-dim, conformal light-front QCD and the 't Hooft Equation of $(1+1)$-dim, large $N_c$ QCD, can be complementary to each other in providing a first approximation to hadron spectroscopy. Together, the two equations play a role in hadronic physics analogous that of the ordinary Schr\"odinger Equation in atomic physics.  
}
\maketitle
\section{Introduction}
\label{intro}
In $(3+1)$-dim light-front QCD, the internal dynamics of a quark-antiquark meson are governed by the Schr\"odinger-like equation: \cite{Brodsky:2014yha,Brodsky:1997de}
\begin{equation}
	\left[ \left(-\frac{\nabla_{b_\perp}^2}{x(1-x)} +\frac{m^2_q}{x} + \frac{m^2_{\bar{q}}}{1-x} \right) + U (x, b_\perp)\right]\Psi(x,\mathbf{b}_\perp)=M^2
	\Psi(x,\mathbf{b}_\perp)	 \;,
	\label{M}
\end{equation}
where $M$ is the meson mass and $\Psi(x,\mathbf{b}_\perp)$ is the meson light-front wavefunction, with $x$ being the light-front momentum fraction carried by the quark, and $\mathbf{b}_\perp$  the transverse distance between the quark and the antiquark. The quark and antiquark masses, $m_q$ and $m_{\bar{q}}$, are effective masses, and $U(x,b_\perp)$ is an effective confining potential. The advantage of formulating QCD on the light-front is that the structure of Eq. \eqref{M} is analogous to that of the ordinary Schr\"odinger Equation, despite Eq. \eqref{M} being fully relativistic and frame-independent. It remains that deriving $U(x,b_\perp)$ from first principles in QCD is an open question. Insights into its analytic form can be gained by introducing a new variable,
\begin{equation}
	\boldsymbol{\zeta}=\sqrt{x(1-x)} \mathbf{b}_\perp 
\end{equation}
and factorizing the wavefunction as
\begin{equation}
	\Psi (x, \zeta, \varphi)= \frac{\phi (\zeta)}{\sqrt{2\pi \zeta}}	 e^{i L \varphi}
 X(x)\,
	\label{full-mesonwf}
\end{equation}
with $L=|L_z^{\mathrm{max}}|$ being the light-front orbital angular momentum, and $X(x)=\sqrt{x(1-x)} \chi(x)$. Assuming that
\begin{equation}
	U(x,\mathbf{\zeta})=U_\parallel(x) + U_\perp(\mathbf{\zeta}) \;,
\end{equation}
Eq. \eqref{M} then separates into
\begin{equation}
	\left[-\frac{d^2}{d \zeta^2}+\frac{4L^2-1}{4 \zeta^2}+U_\perp(\zeta) \right] \phi(\zeta)=M_\perp^2 \phi(\zeta) 
\label{TSE}
\end{equation}
and
\begin{equation}
	\left[\frac{m_q^2}{x}+\frac{m_{\bar{q}}^2}{1-x} +U_\parallel(x) \right] \chi(x)=M_\parallel^2 \chi(x) \;.
\label{LSE}
\end{equation}
with
\begin{equation}
	M^2=M_\perp^2 + M_\parallel^2\,
\end{equation}
and
\begin{equation}
	\int \mathrm{d} x |\chi(x)|^2 = \int \mathrm{d}^2 \boldsymbol{\zeta} |\phi(\zeta)|^2=1 \;.
\end{equation}

As our notation suggests, Eq. \eqref{LSE} governs the longitudinal dynamics while, for a given $x$,  Eq. \eqref{TSE} governs the transverse dynamics. The $x$-dependence of $\zeta$ means that, in this type of factorization, the so-called transverse and longitudinal dynamics do not decouple completely. However, the variable $\zeta$ plays a key role in the search of an analytical form for $U_\perp(\zeta)$ \cite{deTeramond:2005su,deTeramond:2008ht,Brodsky:2006uqa}.

\section{The holographic Schr\"odinger Equation}
\label{LFH}
 If  $\zeta$ is identified with the $5^{\mathrm{th}}$ dimension of $\mathrm{AdS}_5$, and $L^2$ with $(\mu R)^2 + (2-J)^2$, where $\mu$ is a mass parameter and $R$ the radius of curvature in $\mathrm{AdS}_5$, then Eq. \eqref{TSE} becomes the wave equation for the freely propagating spin-$J$ string modes in $\mathrm{AdS}_5$, with the confining potential given by 
\cite{Brodsky:2014yha}
\begin{equation}
	U_{\perp}(\zeta)= \frac{1}{2} \varphi^{\prime \prime}(z) + \frac{1}{4} \varphi^{\prime}(z)^2 + \frac{2J-3}{2 z} \varphi^{\prime}(z) \;,
	\label{dilaton-potential}
	\end{equation}
where $\varphi(z)$ is a dilaton field which distorts the geometry of pure $\mathrm{AdS}_5$. 

At this point, it may seem that the choice for the dilaton field is arbitrary. However, if we insist that the underlying action of  Eq. \eqref{TSE} remains conformally invariant with a mass scale in Eq. \eqref{TSE}, then the latter, which we denote by $\kappa$, can only appear via a harmonic confining potential: $U_\perp(\zeta)=\kappa^4 \zeta^2$. This means that the dilaton field in $\mathrm{AdS}_5$ must be quadratic, $\varphi(z)=\kappa^2 z^2$, and consequently Eq. \eqref{dilaton-potential} implies that 
\begin{equation}
		U_{\perp}(\zeta)=\kappa^4 \zeta^2 + 2 \kappa^2 (J-1)	\;.
	\label{hUeff}	
\end{equation}
Hence Eq. \eqref{hUeff} is fixed simultaneously by the underlying conformal symmetry and  holographic mapping. The conformal algebra leading to the uniqueness of the quadratic potential in Eq. \eqref{hUeff} can be found in \cite{Brodsky:2013ar,Brodsky:2014yha}. Solving the holographic Schr\"odinger Equation, i.e. Eq. \eqref{TSE} with  Eq. \eqref{hUeff}, yields the meson spectrum 
\begin{equation}
	M_{\perp}^2(n_\perp, J, L)= 4\kappa^2 \left( n_\perp + \frac{J+L}{2} \right) 
\label{MT}
\end{equation}
where $n_\perp$ is the principal quantum number, and $J=L+S$ with $S$ being the total quark-antiquark spin. 

The supersymmetric formulation of Eq. \eqref{TSE} allows baryons, mesons and tetraquarks to be considered in a unified framework \cite{Dosch:2015bca,Dosch:2016zdv,Nielsen:2018uyn,Brodsky:2016yod}. This hadronic-level supersymmetry is motivated by an essential feature of colour $\mathrm{SU}(N_c)$: a cluster of $N_c-1$ constituents can be in the same colour representation as the anti-constituent \cite{Brodsky:2016yod}. For $N_c=3$, it means that a diquark (antidiquark) can be in the same colour representation as an antiquark (quark). It follows that the colour confinement dynamics are the same in a quark-diquark baryon, quark-antiquark meson and a diquark-antiquark tetraquark. The supersymmetrization of Eq. \eqref{hUeff} leads to two harmonic-oscillator partner potentials with the same confinement scale, $\kappa$, so that Eq. \eqref{MT}  generalizes to:\cite{Nielsen:2018uyn}
\begin{equation}
	M_{\perp,M}^2=4\kappa^2\left(n_\perp + L_M + \frac{S_M}{2}\right) \;,
	\label{MTM}
\end{equation}
\begin{equation}
	M_{\perp,B}^2=4\kappa^2\left(n_\perp + L_B  + \frac{S_D}{2} + 1\right) \;,
	\label{MTB}
\end{equation}
and
\begin{equation}
	M_{\perp,T}^2=4\kappa^2\left(n_\perp + L_T + \frac{S_T}{2}+1\right) \,,
	\label{MTT}
\end{equation}
where $S_D$ is the lowest value of the total quark-diquark spin in baryons, $S_M$ is the total quark-antiquark spin in mesons, and $S_T$ is total diquark-antidiquark spin in tetraquarks. Eqs. \eqref{MTM}, \eqref{MTB} and \eqref{MTT} show that the masses of baryons with  $L_B=L_M-1$ and $S_D=S_M$ are equal to the masses of mesons with  $L_M$ and $S_M$ and tetraquarks $S_T=S_D$ and $L_T=L_B$. Within a family of superpartners, we can thus write $S\equiv S_D=S_M=S_T$. Note that the lowest lying mesons with $n_\perp=L_M=0$ do not have a baryonic superpartner. Furthermore, pseudoscalar mesons with $n_\perp=L_M=S_M=0$, like the pion, are predicted to be massless.  

In nature, the above superconformal symmetry by broken by non-zero quark masses and longitudinal confinement that generate $M_\parallel$ via Eq. \eqref{LSE}. To solve Eq. \eqref{LSE}, we need to specify the longitudinal confinement potential. Here we take it to be the 't Hooft potential.

\section{The 't Hooft Equation} 
The 't Hooft Equation is derived from the QCD Lagrangian in $(1+1)$-dim, in the large $N_c$ approximation, where only planar diagrams contribute and can be summed to all orders. The result is \cite{tHooft:1974pnl}:
\begin{equation}
\left(\frac{m_q^2}{x}+\frac{m_{\bar{q}}^2}{1-x}\right)\chi(x) +\frac{g^2}{\pi} \mathcal{P} \int {\rm d}y \frac{\chi(x)-\chi(y)}{(x-y)^2}=M^2_\parallel \chi(x) \;,
  \label{tHooft}
\end{equation}
where $\mathcal{P}$ denotes the principal value prescription and $g=g_s \sqrt{N_c}$ is the (finite) 't Hooft coupling with $g_s$ being the strong coupling and $N_c$ the number of colours. Here we shall refer to the `t Hooft coupling as the longitudinal confinement scale. Eq. \eqref{tHooft} coincides with Eq. \eqref{LSE}, if we take
\begin{equation}
	U_\parallel(x)=\frac{g^2}{\pi} \mathcal{P} \int {\rm d}y \frac{\chi(x)-\chi(y)}{(x-y)^2} \;.
\label{UL}
\end{equation}
Fourier transforming Eq. \eqref{UL} to position space, we obtain \cite{Ahmady:2021lsh}
\begin{equation}
	U_L(x^-)= \frac{g^2}{2} P^+|x^-|= g^2 P^+ b_\parallel
	\label{UL-x}
\end{equation}
where $P^+$ is the light-front momentum of the meson and $b_\parallel$ is the longitudinal separation between the quark and the antiquark:  the 't Hooft potential is a linear confinement potential in position space. Note that $\tilde{z}=P^+ x^-$ is the frame invariant longitudinal coordinate introduced in \cite{Miller:2019ysh}. To use the 't Hooft Equation for baryons and tetraquarks,  we transform the antiquark into a diquark (for baryons), followed by the transformation of the quark into an antidiquark (for tetraquarks). Just as $\kappa$, $g$ does not change within a family of superpartners. 

We note that the idea of using the `t Hooft potential as a candidate for $U_\parallel(x)$ was first proposed in \cite{Chabysheva:2012fe}, and that \cite{DeTeramond:2021jnn,Li:2021jqb} use an alternative model \cite{Li:2015zda,Sheckler:2020fbt} for $U_\parallel(x)$, namely
\begin{equation}
	U_\parallel(x)=-\sigma^2 \partial_x (x(1-x)) \partial_x
\label{UL-BFLQ}
\end{equation}
where $\sigma$ is the longitudinal confinement scale. Eq. \eqref{LSE} then admits analytical solutions: \cite{Li:2021jqb}
\begin{equation}
	M_{\parallel}^2= \sigma (m_q + m_{\bar{q}}) (2n_\parallel +1) + \sigma^2 n_\parallel (n_\parallel + 1) + (m_q + m_{\bar{q}})^2 \;.
\label{Mparallel-BLFQ}
\end{equation}
 On the other hand, the `t Hooft Equation needs to be solved numerically and we follow the matrix method outlined in \cite{Chabysheva:2012fe} to do so. We have checked that this numerical method correctly reproduces the analytical solutions, Eq. \eqref{Mparallel-BLFQ}, when using Eq. \eqref{UL-BFLQ}. We draw attention to the recent paper \cite{Weller:2021wog} where a comparison between the different models for the longitudinal potential is carried out.

Previously, the Brodsky-de T\'eramond ansatz \cite{Brodsky:2008pg} was widely used to accommodate light quark masses while neglecting longitudinal confinement. Despite its phenomenological success, the shortcoming of the Brodsky-de T\'eramond ansatz is that it is not consistent \cite{Li:2021jqb} with the Gell-Mann-Oakes-Renner (GMOR) relation \cite{PhysRev.175.2195} . Another ansatz designed to agree with the GMOR relation can be found in \cite{Gutsche:2012ez}.  

\section{Symmetry constraints}

Chiral symmetry breaking in QCD is encapsulated in the GMOR relation which predicts that $M_\pi^2 \propto m_{u/d}$ in the limit of small quark masses. Now the holographic Schr\"odinger Equation predicts that $M_\pi=0$ while the `t Hooft Equation predicts that, if $m_{u/d} \ll g$, then  $M_\pi^2 \propto g m_{u/d}$.  Hence, the two equations together correctly satisfy the GMOR relation, provided that $g$ does not scale with $m_{u/d}$ in the chiral limit.     
 
 Heavy Quark Effective Theory (HQET) predicts that the mass difference between heavy-light vector and pseudoscalar mesons in their ground states, is suppressed by the heavy quark mass: $M_{qQ}^V - M_{qQ}^P \sim 1/m_Q$. Now, the holographic Schr\"odinger Equation predicts that, if $m_Q \gg \kappa$, $M^V_{qQ} \sim m_Q(1+\kappa^2/2m_Q^2)$ and that $M_{qQ}^P=0$, while the `t Hooft Equation predicts that $M_{qQ}^{P/V} \sim m_Q$ if $m_Q \gg g$. Hence,  together, two equations satisfy the HQET constraint, provided that $\kappa$ does not scale with $m_Q$ in the heavy quark limit.  
  
 Rotational symmetry is restored in the centre-of-mass (CM) frame of a heavy-heavy meson since the system is essentially non-relativistic. In the non-relativistic limit, the light-front holographic potential differs from the `t Hooft potential: $U_\perp \sim (\kappa^4/4) b_\perp^2$ (since $x \sim 1/2$) and $U_\parallel \sim 2m_Q b_\parallel$ (since $P^+=M \sim 2m_Q$). However, the rotational symmetry constraint applies to instant form potentials in the CM frame. Using the relation between light-front and instant-form potentials given in \cite{Trawinski:2014msa}, we find that  both light-front potentials correspond to linear instant-form potentials in the CM frame: $V_\perp \sim(\kappa^2/2) b_\perp$ and $V_\parallel \sim (g^2/2) b_\parallel$. Rotational symmetry implies that $V_\perp =V_\parallel$, and is satisfied provided that $g = \kappa$ in the non-relativistic limit.

The above constraints motivate us to consider $g$ to be independent of $m_{u/d}$ in light mesons, $\kappa$ to be  independent of $m_Q$ in heavy-light mesons, and $g$ to coincide with $\kappa$ in heavy-heavy mesons. These statements also apply to their baryon and tetraquark superpartners since the confinement scales do not change within a family of superpartners. 


\section{Computing the hadronic spectrum}
When solving the 't Hooft Equation, an additional quantum number, $n_\parallel$, emerges. Thus, each hadronic state is identified by four quantum numbers, $n_\perp, L, S, n_\parallel$, with their squared masses given by 
\begin{equation}
	M_{M}^2=M^2_{\perp,M}(n_\perp, L_M, S_M, \kappa) + M^2_{\parallel,M}(n_\parallel, m_q, m_{\bar{q}}, g)\;,
	\label{MM}
\end{equation}
\begin{equation}
	M_{B}^2=M^2_{\perp,B}(n_\perp, L_B, S_D, \kappa) + M^2_{\parallel,B}(n_\parallel, m_q, m_{[qq]}, g) \;,
	\label{MB}
\end{equation}
\begin{equation}
	M_{T}^2= M^2_{\perp,B}(n_\perp, L_T, S_T, \kappa) + M^2_{\parallel,T}(n_\parallel, m_{[\bar{q}\bar{q}]}, m_{[qq]}, g)\,,	
	\label{MT}
\end{equation}
while their parity and charge conjugation are given by
\begin{equation}
	P=(-1)^{L_M+1}=(-1)^{L_B}=(-1)^{L_T}\;.
\label{P}
\end{equation}
and
\begin{equation}
	C=(-1)^{n_\parallel + L_M+S_M}=(-1)^{n_\parallel + L_T + S_T-1} \;.
\label{C}
\end{equation}
\`A posteriori, we find that 
\begin{equation}
	n_{\parallel} \ge n_{\perp} + L 
	\end{equation}
i.e. an orbital and/or radial excitation in the transverse dynamics is always accompanied by an excitation in the longitudinal dynamics. 

To calculate the hadronic spectrum, we need to specify the numerical values of the confinement scales and quark masses. We take the diquark mass to be simply twice the quark mass and we use the universal value of the transverse confinement scale, $\kappa=0.523$ GeV \cite{Brodsky:2016rvj}. Our choices for longitudinal confinement scale, $g$, and the quark masses are shown in Table ~\ref{tab-1}. Note that we find it necessary to allow $g$ to vary with the number of heavy quarks in the hadron, while remaining the same within a family of superpartners and coinciding with $\kappa$ for hadrons with two heavy quarks. The quark masses in the light hadrons are identical to those used previously in light-front holography \cite{Brodsky:2014yha}. 

\begin{table}
\centering
\caption{Quark masses and longitudinal confinement scale in $\mathrm{GeV}$. Note that we use $\kappa=0.523$ GeV for all hadrons, with the constraint $g=\kappa$ for hadrons with two heavy quarks.}
\label{tab-1}       
\begin{tabular}{llllll}
\hline
Hadron & $g$ & $m_{u/d}$ & $m_s$ & $m_c$ & $m_b$  \\ \hline
Light~ & 0.128 & 0.046 & 0.357 & - & - \\
 Heavy-light & 0.410 & 0.330 & 0.500 & 1.370 & 4.640 \\
 Heavy-heavy~ & 0.523  & - & - & 1.370 & 4.640  \\ \hline
\end{tabular}
\end{table}

To illustrate qualitatively the agreement with spectroscopic data, we show a sample of Regge trajectories for hadrons with no, one or two charm quarks, taken from \cite{Ahmady:2021yzh} where the full sets of numerical results can be found. In general, the agreement for mesons and baryons is quite good. For the tetraquark candidates, which we identify following \cite{Nielsen:2018uyn},  there is a couple of very notable disagreements. These discrepancies were already present when longitudinal dynamics were neglected and are thus not alleviated by accounting for the latter. 

\begin{figure}[http]
		\includegraphics[width=7cm,clip]{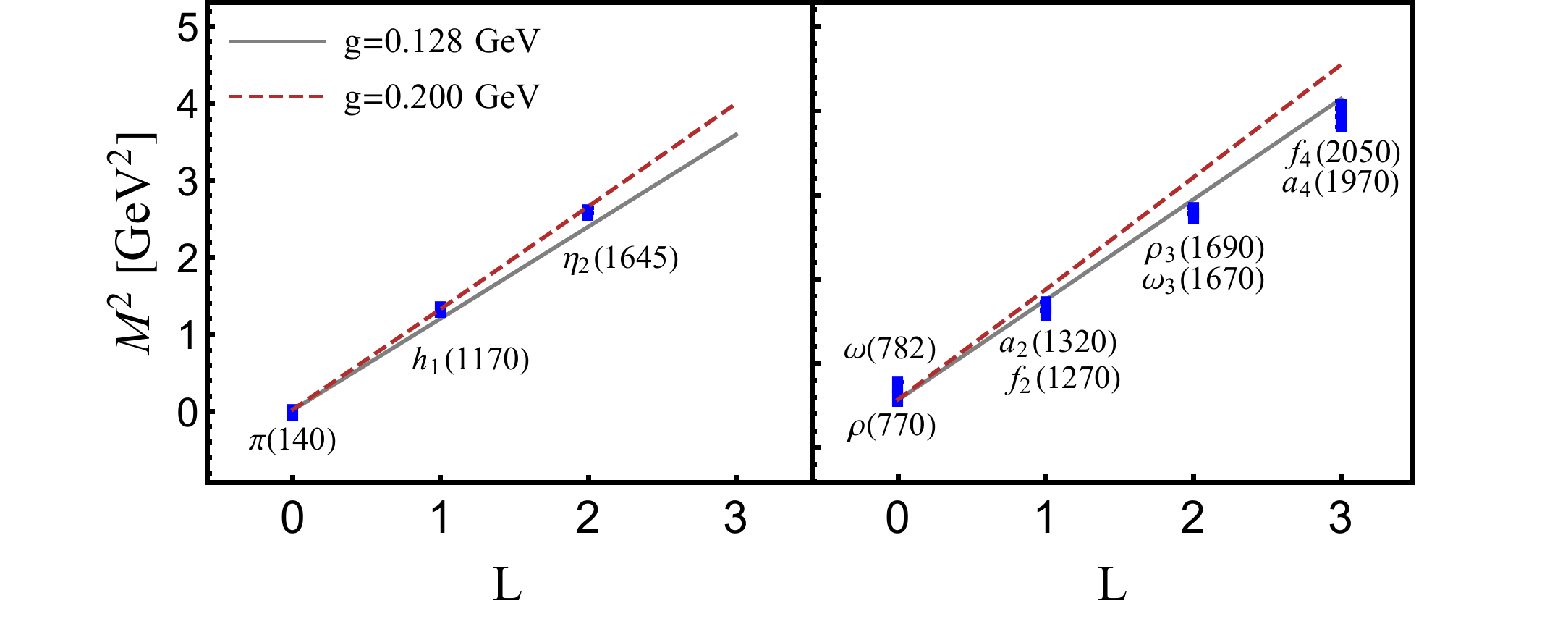}\includegraphics[width=7cm,clip]{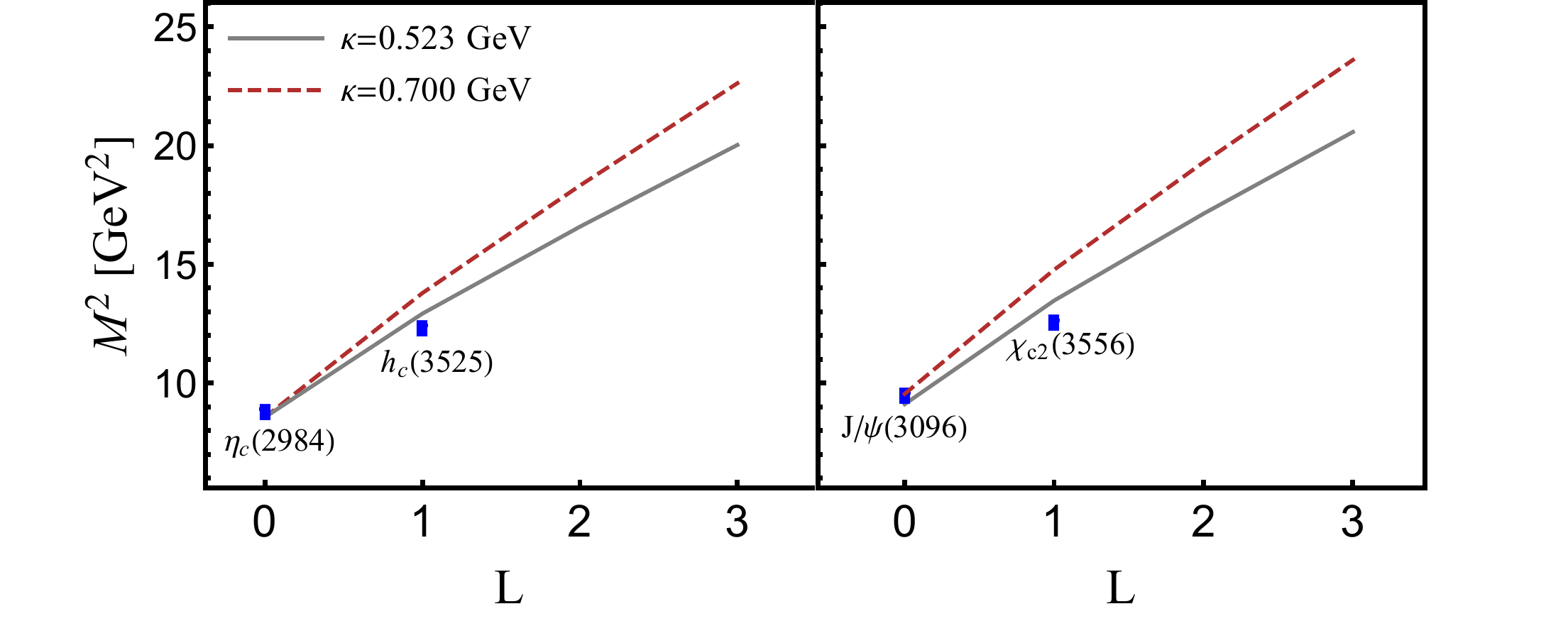}		\caption{Left: Sensitivity of the Regge trajectories for $\pi$ and $\rho$ families to $g$: $g=0.128$ GeV (solid-gray) and $g=0.200$ GeV (dashed-dark-red). Right: Sensitivity of Regge trajectories for $\eta_c$ and $J/\psi$ families to $\kappa$: $\kappa=0.523$ GeV (solid-gray) and $\kappa=0.700$ GeV (dashed-dark-red).}
\label{diff-g-light}
\end{figure}

Given that, except for the pion and the kaon, $M_\perp \ll M_\parallel$ for light hadrons and that $M_\parallel \gg M_\perp$ in heavy hadrons, it is legitimate to ask  whether the Regge trajectories are sensitive to $g$ for light hadrons and to $\kappa$ for heavy hadrons. The answer is yes, as illustrated by Fig. \ref{diff-g-light}. This means that the statement that $\kappa$ remains universal across the full spectrum is not a trivial one.

\begin{figure}[h]
\centering
\includegraphics[width=10cm,clip]{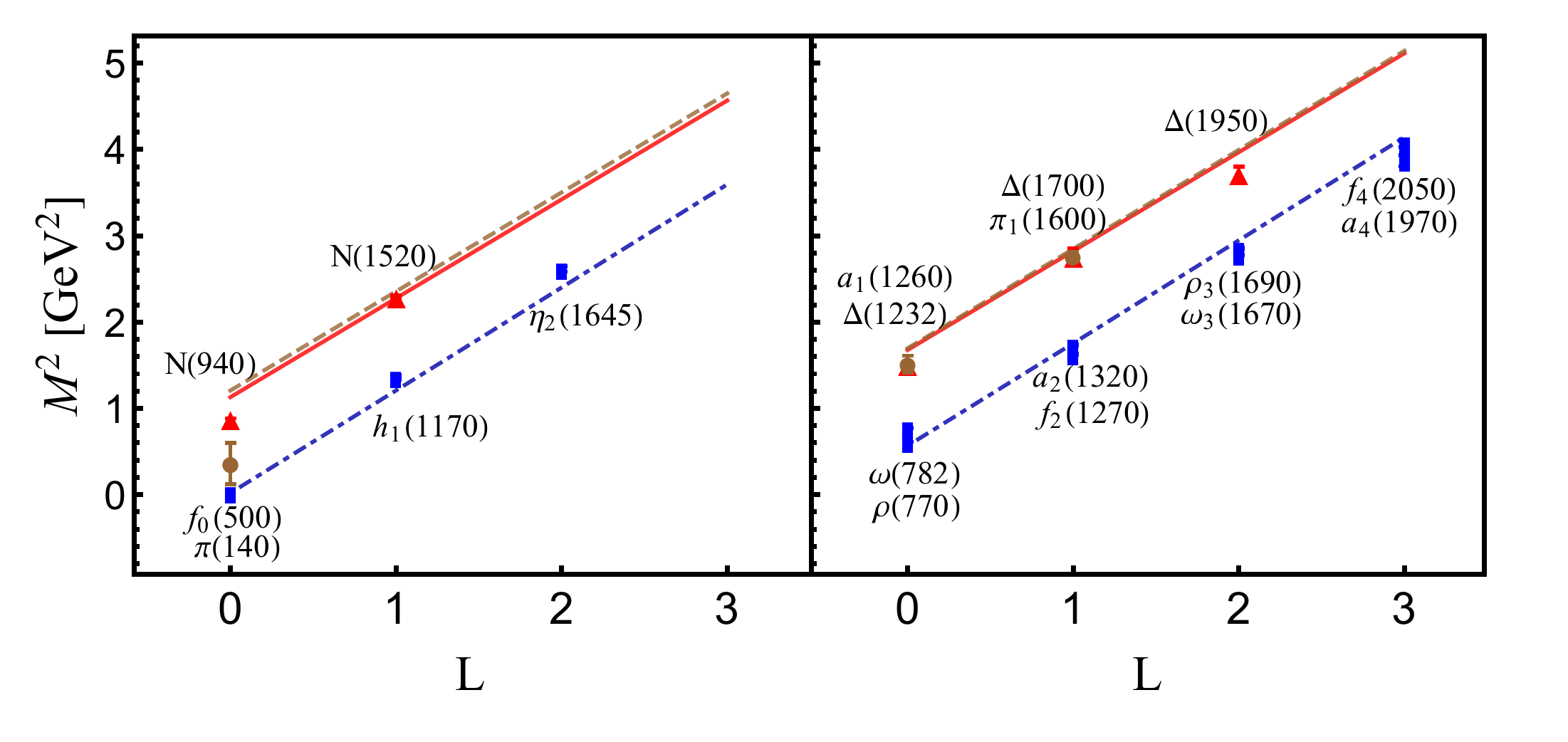}
\caption{Regge trajectories for the light hadrons. Dot-dashed-blue: mesons, solid-red: baryons, dashed-brown: tetraquarks. Data from PDG \cite{Zyla:2020zbs}.}
\label{fig-1}       
\end{figure}


\begin{figure}[h]
\centering
\includegraphics[width=10cm,clip]{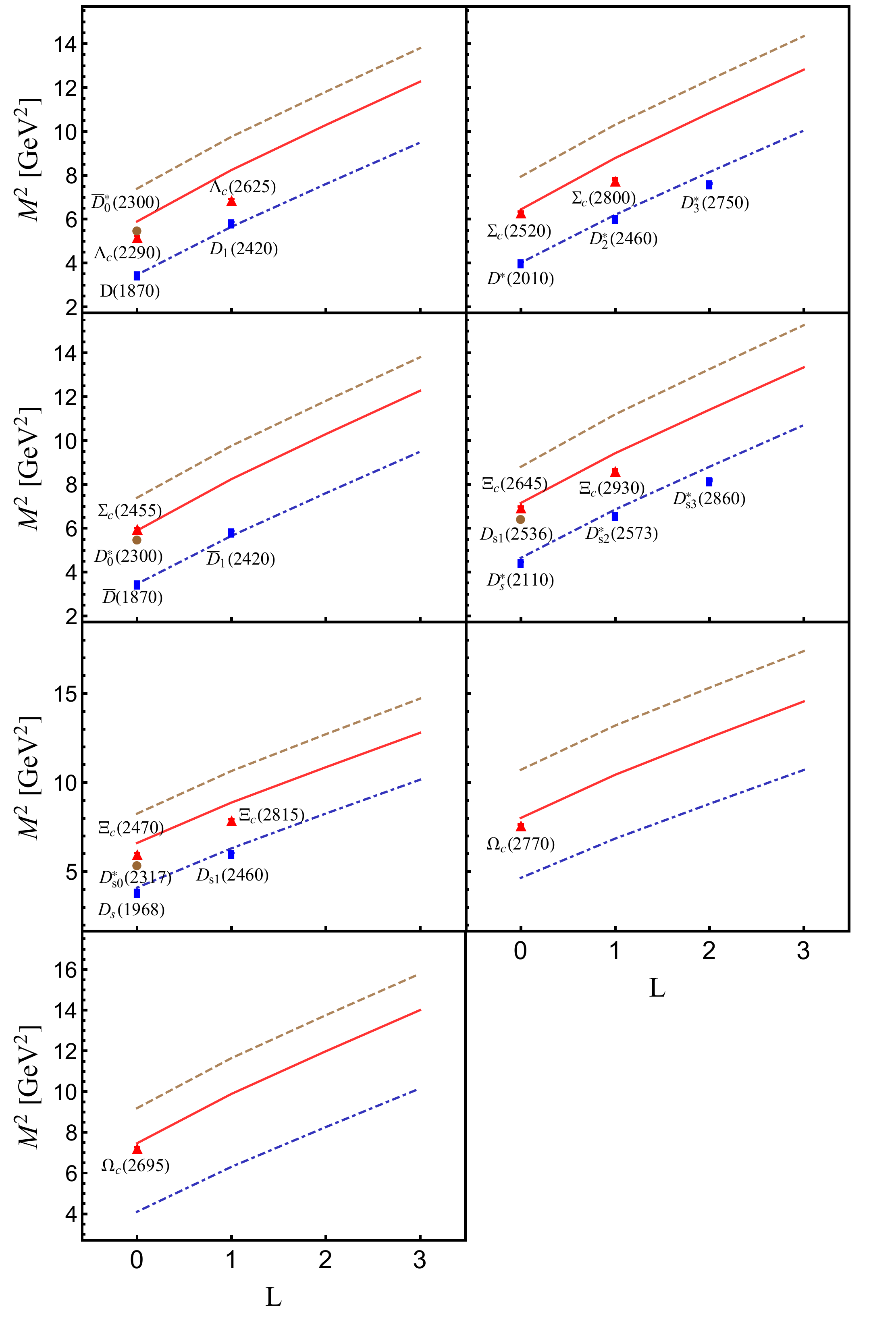}
\caption{Regge trajectories for hadrons with one charm quark. Dot-dashed-blue: mesons, solid-red: baryons, dashed-brown: tetraquarks. Data from PDG \cite{Zyla:2020zbs}.}
\label{fig-3}       
\end{figure}

\begin{figure}[h]
\centering
\includegraphics[width=10cm,clip]{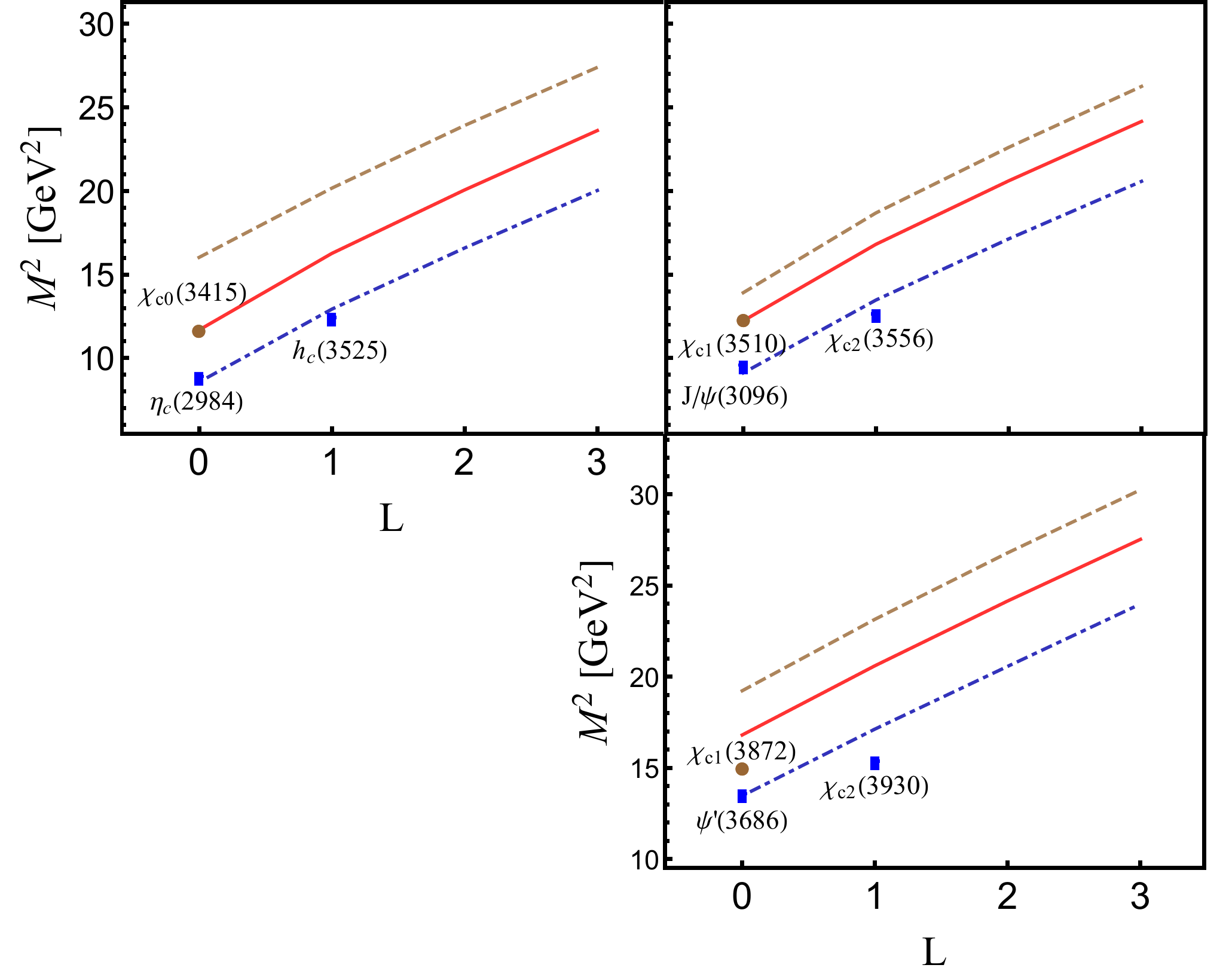}
\caption{Regge trajectories for hadrons with two charm quarks. Dot-dashed-blue: mesons, solid-red: baryons, dashed-brown: tetraquarks. Data from PDG \cite{Zyla:2020zbs}.}
\label{fig-4}       
\end{figure}

\section{Conclusions}
We have argued that, together, the holographic Schr\"odinger Equation and the 't Hooft Equation describe the main features of the full hadron spectrum. The confinement scale of the holographic Schr\"odinger Equation is universal across the full spectrum. On the other hand, the confinement scale of the `t Hooft Equation  depends on the number of heavy quarks in the hadrons, while remaining the same within a family of superpartners and coinciding with the holographic scale in hadrons with two heavy quarks.

\section{Acknowledgements}
I thank the organizers of the Virtual Tribute to Confinement and the Hadron Spectrum at the University of Stavanger, Norway, for a successful conference. I am grateful to my coauthors of \cite{Ahmady:2021lsh,Ahmady:2021yzh} on which this talk is based. I  also thank Stan Brodsky, Guy de T\'eramond and Gerald Miller for useful discussions. This research is supported  by an Individual Discovery Grant (SAPIN-2020-00051) from the Natural Sciences and Engineering Research Council of Canada (NSERC).

%
\bibliography{Sandapen1.bib}
%
%
%
%

\end{document}